\documentclass[11pt]{article}
\usepackage{amsmath}
\usepackage{amsfonts}
\usepackage{amssymb}
\usepackage{graphicx}

\def\bea{\begin{eqnarray}}
\def\eea{\end{eqnarray}}
\title{  AdS/BCFT in non-relativistic limit }

\date{}

\begin{document}
\begin{center}
\LARGE { \bf  Anti--de Sitter/ boundary conformal field theory
correspondence in the non-relativistic limit

  }
\end{center}
\begin{center}
{\bf M. R. Setare\footnote{rezakord@ipm.ir} \\  V. Kamali\footnote{vkamali1362@gmail.com}}\\
 { Department of Science, Payame Noor University, Bijar, Iran}
 \\
 \end{center}
\vskip 3cm

\begin{abstract}
Boundary conformal field theory (BCFT) is the study of conformal
field theory (CFT) in semi-infinite space-time. In non-relativistic
limit ($x\rightarrow\epsilon x, t\rightarrow t, \epsilon\rightarrow
0$), boundary conformal algebra changes to boundary Galilean
conformal algebra (BGCA). In this work, some aspects of AdS/BCFT in
non-relatvistic limit were  explored. We constrain correlation
functions of Galilean conformal invariant fields with BGCA
generators. For a situation with a boundary condition at surface
$x=0$ ($z=\overline{z}$), our result is agree with non-relativistic
limit of BCFT two-point function. We also, introduce holographic
dual of boundary Galilean conformal field theory.

\end{abstract}

\newpage

\section{Introduction}
 Recently, there has been some
interest in extending the AdS/CFT correspondence to
non-relativistic field theories \cite{Luk, 2}, where the
non-relativistic conformal symmetry was obtained by a parametric
contraction of the relativistic conformal group. Galilean
conformal algebra (GCA) arises as a contraction relativistic
conformal algebras \cite{Luk,co,Du}, where in $d=4$ the Galilean
conformal group is a fifteen parameter group which contains the
ten parameter Galilean subgroup. Beside Galilean conformal
algebra, there is another non-relativistic algebra, the twelve
parameter Schr\"{o}dinger algebra \cite{so,bal}. The dilatation
generator in the Schr\"{o}dinger group scales space and time
differently, $x_i\rightarrow \lambda x_i$, $t\rightarrow
\lambda^2 t$, but in contrast the corresponding generator in GCA
scales space and time in the same way, $x_i\rightarrow \lambda
x_i$, $t\rightarrow \lambda t$. Infinite dimensional  Galilean
conformal group has been reported  in \cite{co}. The generators
of this group are : ~$L^n=-(n+1)t^n
x_i\partial_i-t^{n+1}\partial_t$,~$M^{n}_i=t^{n+1}\partial_i$
and~$J_{ij}^n=-t^n(x_i\partial_j-x_j\partial_i)$~ for an arbitrary
integer $n$, where  $i$ and  $j$ are specified by the spatial
directions which obey commutation relation of the
Virasoro-Kac-Moody algebra \cite{co}. There is a finite
dimensional subgroup of the infinite dimensional Galilean
conformal group which is generated by ~($J_{ij}^0, L^{\pm 1},
L^0, M_i^{\pm 1}, M_i^0$). These generators are obtained by
contraction ($~t\rightarrow t, ~x_i \rightarrow \epsilon  x_i,
~\epsilon \rightarrow 0,~ v_i \sim \epsilon$ ) of the relativistic
conformal generators. The gravity dual of finite GCA was
considered in \cite{co,Du,va} and the metric with finite 2d GCA
isometry was obtained in \cite{ku}.\\
The presence of free surfaces or walls in macroscopic systems
which are at the critical point, lead to the large variety of
physical effects. Since, using boundary condition effect is shown
to be very helpful in various branch in physics, the systems with
boundary conditions have been considered by both theorists
\cite{hen} and experimentalists \cite{bin}. The situation with
walls or free surfaces opens a new area in condensed matter
physics \cite{dom}. In reference \cite{jans}, the research on
semi-infinite systems which exhibits a non-equilibrium bulk phase
transitions was initiated and  the effects of boundary condition
on direct percolation were
considered.\\
Holographic dual of a conformal field theory with a boundary (BCFT)
was proposed in \cite{ta}.  The main idea of AdS/BCFT correspondence
was started with asymptotically AdS geometry with Neumann boundary
condition on the metric as one approaches to the boundary
\cite{ta,fu}. The geometry is modified by imposing two different
boundary conditions on the metric. The boundary is divided  into two
parts $\partial M= N \bigcup Q$ where $\partial Q=\partial N$
\cite{ta}. The metric has Neumann boundary condition on $Q$ and
Dirichlet boundary condition on $N$. With this boundary condition
the AdS geometry is divided into two parts and the gravitational
theory lives in one part of this space. This modified geometry could
provide a holographic dual for BCFT. Boundary conformal field theory
(BCFT) is defined in domains  with a boundary \cite {ca}. In this work
we extend AdS/BCFT correspondence to non-relativistic version. When
non-relativistic CFT lives in semi-infinite space, one sector of
Galilean conformal group is removed. For example, if we have a
boundary condition on surface $z=\overline{z}~( t-x=t+x)$ or $x=0$,
translation, boost and spatial-special conformal transformations are
removed. So, two-point function in this situation is completely
different from situation without boundary condition (free space).
Two-point function of BCFT in the situation with a boundary
condition at surface $z=\overline{z}$ was calculated in
\cite{al,se}. In this paper we calculate two-point and three point
functions of BGCA from gravity dual \cite{va} and quantum field
theory method in the boundary \cite{1,ba}. Our results agree with
results \cite{al,se} in non-relativistic limit. We also, introduce
holographic dual of non-relativistic limit of BCFT (BGCA). The paper
organized as follow: In section 2 we give a brief review of 2d GCA.
In section 3 we calculate two-point and three point correlation
functions of Galilean conformal invariant fields in semi-infinite
space. In section 4 we introduce holographic dual of
non-relativistic BCFT, then we calculate two-point function from
gravity dual. Finally, in section 5, we close by some concluding
remarks.
\section{GCA in 2d}
Galilean conformal algebra (GCA) in 2d is obtained by contracting
2d conformal symmetry \cite{1}. Two-dimensional  Conformal algebra
 is described by two copies of Virasoro
algebra. In quantum field theory (QFT) level, two-dimensional
($z=t+x$, $\overline{z}=t-x$) CFT generators
\begin{eqnarray}\label{1}
\mathcal{L}_{n}=z^{n+1}\partial_{z},
~~~~~~~~~~~~~~~~~~~~~~~~\overline{\mathcal{L}}_n=\overline{z}^{n+1}\partial_{\overline{z}},
\end{eqnarray}
obey the  Virasoro algebra
\begin{eqnarray}\label{2}
[\mathcal{L}_m,\mathcal{L}_n]=(n-m)\mathcal{L}_{m+n}+\frac{c_{R}}{12}m(1-m^2)\delta_{m+n,0}, \\
\nonumber
[\overline{\mathcal{L}}_m,\overline{\mathcal{L}}_n]=(n-m)\overline{\mathcal{L}}_{m+n}+\frac{c_{L}}{12}m(1-m^2)\delta_{m+n,0}.
\end{eqnarray}
In non-relativistic limit ($t\rightarrow t$, $x\rightarrow\epsilon
x$ with $\epsilon\rightarrow 0$),  the GCA generators $L_{n}$ and
$M_n$ are constructed by Virasoro generators
\begin{eqnarray}\label{3}
L_n=\lim_{\epsilon\rightarrow
0}(\mathcal{L}_n+\overline{\mathcal{L}}_n)=(n+1)t^n\partial_{x}+t^{n+1}\partial_x, ~~~~~~~~\\
\nonumber M_{n}=-\lim_{\epsilon \rightarrow 0 }\epsilon
(\mathcal{L}_n-\overline{\mathcal{L}}_n)=-t^{n+1}\partial_{x}.
~~~~~~~~~~~~~~~~~~~~~
\end{eqnarray}
 From Eqs.(\ref{2}) and (\ref{3}), one obtains centrally extended 2d GCA
\begin{eqnarray}\label{4}
[L_m,L_n]=(n-m)L_{m+n}+C_1m(1-m^2)\delta_{m+n,0}, ~\\
\nonumber [L_m,M_n]=(n-m)M_{m+n}+C_2m(1-m^2)\delta_{m+n,0}, \\
\nonumber [M_n,M_m]=0.
~~~~~~~~~~~~~~~~~~~~~~~~~~~~~~~~~~~~~~~~~~~~~
\end{eqnarray}
The GCA central charges ($C_1$, $C_2$) are related to CFT central
charges ($c_L$, $c_R$) as:
\begin{eqnarray}\label{5}
C_1=\lim_{\epsilon\rightarrow 0}\frac{c_L+c_R}{12},
~~~~~~C_2=\lim_{\epsilon\rightarrow
0}(\epsilon\frac{c_L-c_R}{12}).
\end{eqnarray}
From above equations, for a non-zero  and finite ($C_2$, $C_1$) in
the limit $\epsilon\rightarrow 0$, it can be seen that we need
$c_L-c_R\propto O(\frac{1}{\epsilon})$ and $c_L+c_R \propto
O(1)$.  Similarly, rapidity $\xi$ and scaling dimensions
$\Delta$, which are the eigenvalues of $M_0$ and $L_0$
respectively, are given by
\begin{eqnarray}\label{6}
\Delta=\lim_{\epsilon \rightarrow 0}(h+\overline{h}),
~~~~~~~~~\xi=-\lim_{\epsilon\rightarrow 0}\epsilon
(h-\overline{h}),
\end{eqnarray}
where $h$ and $\overline{h}$ are eigenvalues of $\mathcal{L}_0$
and $\overline{\mathcal{L}}_0$ respectively. Equation (\ref{6})
tells us that, $h+\overline{h}$ is of order $O(1)$ while
$h-\overline{h}$ must be order $O(\frac{1}{\epsilon})$, for the
finite $\Delta$, $\xi$.
\section{ Correlation functions in semi-infinite space}
In this section we  find the correlation functions in semi-infinite
space with a boundary condition at surface $z=\overline{z}$. We
now turn to derive the consequences of Galilean  conformal
invariance for the correlation. In general, we expect a
quasi-primary field  $\mathcal{O}$   to be characterized by its
Galilean conformal dimension $\Delta$ and rapidity $\xi$ (these
fields are invariant under finite sub-group that is generated by
sub-algebra \{$L_{-1},M_{-1}, L_0$,$M_{0}, L_1,M_1$\}). We would
like to find the form of two-point and three-point functions of the  Galilean
conformal invariant operators in semi-infinite space. Firstly,
we  find the form of the commutators $[\mathcal{L}_n,
\mathcal{O}]$ and $[\overline{\mathcal{L}}_n, \mathcal{O}]$, then
we obtain  the form of $[L_n,\mathcal{O}]$ and $[M_n,\mathcal{O}]$
as following
\begin{eqnarray}\label{7}
[\mathcal{L}_n,\mathcal{O}(z,\overline{z})]=[\mathcal{L}_n,U\mathcal{O}(0)U^{-1}]=[\mathcal{L}_n,U]\mathcal{O}(0)U^{-1}+U\mathcal{O}(0)[\mathcal{L}_n,U^{-1}]\\
\nonumber+U[\mathcal{L}_n,\mathcal{O}(0)]U^{-1}=
U\{U^{-1}\mathcal{L}_nU-\mathcal{L}_n\}\mathcal{O}(0)U^{-1}~~~~~~~~~~~~~~~~~~~~~~~~~\\
\nonumber +U\mathcal{O}(0)\{\mathcal{L}_n-U^{-1}\mathcal{L}_n
U\}U^{-1}+\delta_{n,0}h\mathcal{O}(z,\overline{z})~~~~~~~~~~~~~~~~~~~~~~~~
  \end{eqnarray}
$U$ and $\mathcal{O}(z,\overline{z})$ are defined as
\begin{eqnarray}\label{8}
\mathcal{O}(z,\overline{z})=U\mathcal{O}(0)U^{-1}~~~~~where~~~~~~~U=e^{z\mathcal{L}_{-1}+\overline{z}\overline{\mathcal{L}}_{-1}}
  \end{eqnarray}
By using the Hausdorff formula we get

\begin{eqnarray}\label{9}
U^{-1}\mathcal{L}_nU=e^{-z\mathcal{L}_{-1}-\overline{z}\overline{\mathcal{L}}_{-1}}\mathcal{L}_n
e^{z\mathcal{L}_{-1}+\overline{z}\overline{\mathcal{L}}_{-1}}=e^{-z\mathcal{L}_{-1}}\mathcal{L}_n
e^{z\mathcal{L}_{-1}} \\
 \nonumber
=\mathcal{L}_n+[\mathcal{L}_n,z\mathcal{L}_{-1}]+\frac{1}{2!}[[\mathcal{L}_n,z\mathcal{L}_{-1}],z\mathcal{L}_{-1}]+...\\
\nonumber= \sum_{k=0}^{n+1}\frac{(n+1)!}{(n+1-k)!k!}(z)^k
\mathcal{L}_{n-k}~~~~~~~~~~~~~~~~~~~~
  \end{eqnarray}
and
\begin{eqnarray}\label{10}
\mathcal{L}_{n}^{'}=U^{-1}\mathcal{L}_nU-\mathcal{L}_n=\sum_{k=1}^{n+1}\frac{(n+1)!}{(n+1-k)!k!}(z)^k
\mathcal{L}_{n-k}
  \end{eqnarray}
The Eq.(\ref{7}) gives us
\begin{eqnarray}\label{11}
 [\mathcal{L}_n,\mathcal{O}(z,\overline{z})]=U\{[\mathcal{L}_n^{'},\mathcal{O}(0)]+\delta_{n,0}h\mathcal{O}(0)\}U^{-1}~~~~~~~~~\\ \nonumber
 =z^{n+1}[\mathcal{L}_{-1},\mathcal{O}(z,\overline{z})]+z^n(n+1)U[\mathcal{L}_0,\mathcal{O}(0)]U^{-1}~~~~~~
\end{eqnarray}
Now we have $[\mathcal{L}_{-1},\mathcal{O}]=\partial_{z}\mathcal{O}$~  and~
$[\mathcal{L}_0,\mathcal{O}]=h\mathcal{O}$ ($\mathcal{L}_0$~and~$\mathcal{L}_{-1}$ generate
$z$-dilatation and $z$-translation, respectively). Hence we obtain
(for $n\geq-1$)
\begin{eqnarray}\label{12}
 [\mathcal{L}_n,\mathcal{O}(z,\overline{z})]=(z^{n+1}\partial_{z}+(n+1)hz^n)\mathcal{O}
  \end{eqnarray}
We can exchange $\mathcal{L}_n$ with  $\overline{\mathcal{L}}_n$ and  using the above
steps (\ref{7})-(\ref{11}). We get
\begin{eqnarray}\label{13}
[\overline{\mathcal{L}}_n,\mathcal{O}(z,\overline{z})]=(\overline{z}^{n+1}\partial_{\overline{z}}+(n+1)h\overline{z}^n)\mathcal{O}
\end{eqnarray}
  From Eqs.(\ref{12}),(\ref{13}) and by using the definitions of
  $L_n$ and $M_n$ (\ref{3}), we can find the form of commutators $[L_n,\mathcal{O}]$ and $[M_n,\mathcal{O}]$
\begin{eqnarray}\label{14}
[L_n,\mathcal{O}]=(t^{n+1}\partial_t+(n+1)t^nx\partial_x+(n+1)(t^n\Delta-nxt^{n-1}\xi))\mathcal{O}\\
\nonumber
[M_n,\mathcal{O}]=(t^{n+1}\partial_x-(n+1)t^n\xi)\mathcal{O}~~~~~~~~~~~~~~~~~~~~~~~~~~~~~~~~~~~~~
\end{eqnarray}
Correlation  functions of GCA are constrained  by the above equations
in free space \cite{ba}. If we have a boundary in $x$ direction,
symmetries in this direction is removed obviously. So, in the
situation with a boundary condition at surface $x=0$
($z=\overline{z}$), Galilean symmetry group reduces to one copy
of non-relativistic version of Virasoro group which is generated
by $L_n$ \cite{al,se}. We can use this subgroup to calculate
two-point function. Firstly, we consider the invariance under time
translation which is generated by $L_{-1}$
\begin{eqnarray}\label{15}
<0\mid[L_{-1},G]\mid0>=0~~\Rightarrow~~
G=G(x_1,x_2,\tau)~~~~~~\tau=t_1-t_2
\end{eqnarray}
where $G=<\mathcal{O}_1\mathcal{O}_2>$ is two-point function of
two quasi-primary operators $\mathcal{O}_1$ and $\mathcal{O}_2$.
Invariance under dilatation constrains two-point function as
\begin{eqnarray}\label{16}
<0\mid[L_0,G]\mid0>=0~~~~~~~~~~~~~~~~~~~~~~~~~~~~~~~~~~~~~~~~~\\
\nonumber \Rightarrow
\sum_{i=1}^2(t_i\partial_{t_i}+x_i\partial_{x_i}+\Delta_i)G=0~~~~~~~~~~~~~~~~~~~~~~~~~~~~\\
\nonumber
(\tau\partial_{\tau}+x_1\partial_{x_1}+x_2\partial_{x_i}+\Delta)G=0~~~~~~~~~\Delta=\Delta_1+\Delta_2
\end{eqnarray}
Invariance under spatial component of special conformal
transformation is
\begin{eqnarray}\label{17}
<0\mid[L_1,G]\mid0>=0~~~~~~~~~~~~~~~~~~~~~~~~~~~~~~~~~~~~~~~~~~~~~\\
\nonumber \Rightarrow
\sum_{i=1}^{2}(t_{i}^{2}\partial_{t_{i}}+2t_{i}x_{i}\partial_{x_i}+2t_{i}\Delta_{i}-x_{i}\xi_{i})G~~~~~~~~~~~~~~~~~~~~~~~~~~~\\
\nonumber
= ((t_{1}^2-t_{2}^2)\partial_{\tau}+2(t_{1}x_{1}\partial_{x_{1}}+t_{2}x_{2}\partial_{x_{2}})~~~~~~~~~~~~~~~~~~~~~~~~\\
\nonumber
+2(t_{1}\Delta_{1}+t_{2}\Delta_{2}-x_{1}\xi_{1}-x_{2}\xi_{2}))G~~~~~~~~~~~~~~~~~~~~~~~~~~\\
\nonumber
=(\tau^2\partial_{\tau}+2t_{2}(\tau\partial_{\tau}+x_{1}\partial_{x_{1}}+x_{2}\partial_{x_{2}})~~~~~~~~~~~~~~~~~~~~~~~~~~~\\
\nonumber -2(x_1\xi_1+x_2\xi_1)+2\tau x_1\partial_{x_{2}}
+2(t_1\Delta_1+t_2\Delta_2))G~~~~~~~~\\
\nonumber =(\tau^2\partial_{\tau}-2(x_1\xi_1+x_2\xi_1)+2\tau
x_1\partial_{x_{2}}+2\tau\Delta_{1})G=0~~~~~~~
\end{eqnarray}
where in the last step, Eq.(\ref{16}) was used. We make the
following ansatz
\begin{eqnarray}\label{18}
G(x_1,x_2,\tau)=\tau^{-2\Delta_1}G'(u,v),~~~~~~~u=\frac{x_1}{\tau},~~~~~~~~v=\frac{x_2}{\tau}
\end{eqnarray}
so, Eq. (\ref{17}) gives
\begin{equation}\label{19}
(u\partial_{u}-v\partial_{v}-2(u\xi_1+v\xi_2))G'(u,v)=0,
\end{equation}
Solution of this equation is  \cite{hen,ka}
\begin{equation}\label{20}
G'(u,v)= \chi(uv)\exp(2(u\xi_1-v\xi_2))
\end{equation}
where $\chi$ is an arbitrary function.
The final result for two-point function is
\begin{eqnarray}\label{21}
G(x_1,x_2,\tau)=\delta_{\Delta_1,\Delta_2}\tau^{-2\Delta}\chi(\frac{x_1x_2}{\tau^2})\exp(\frac{2}{\tau}(x_1\xi_1-x_2\xi_2))
\end{eqnarray}
where $\Delta=\Delta_1=\Delta_2$. It is clear  that, two-point
function  near the boundary  is different  from other places
\cite{ba}. Two-point function of BCFT for scalar fields was
calculated in \cite{al,se}
\begin{eqnarray}\label{22}
G(z_1,z_2,\overline{z}_1,\overline{z}_2)=\frac{1}{4}(\frac{1}{\mid z_1-z_2\mid^{2\Delta}}+\frac{1}{\mid \overline{z}_1-\overline{z}_2\mid^{2\Delta}}\\
\nonumber +\frac{1}{\mid
z_1-\overline{z}_2\mid^{2\Delta}}+\frac{1}{\mid
\overline{z}_1-z_2\mid^{2\Delta}})
\end{eqnarray}
In non-relativistic limit  ($t\rightarrow t,x\rightarrow \epsilon
x$) we have
\begin{eqnarray}\label{}
\lim_{\epsilon\rightarrow 0}(z_1-z_2)=\lim_{\epsilon\rightarrow
0}(t_1+\epsilon x_1-t_2-\epsilon x_2)=t_1-t_2\\
\nonumber \lim_{\epsilon\rightarrow
0}(\overline{z}_1-\overline{z}_2)=\lim_{\epsilon\rightarrow
0}(t_1-\epsilon x_1-t_2+\epsilon x_2)=t_1-t_2\\
\nonumber \lim_{\epsilon\rightarrow
0}(z_1-\overline{z}_2)=\lim_{\epsilon\rightarrow 0}(t_1+\epsilon
x_1-t_2+\epsilon x_2)=t_1-t_2\\
\nonumber \lim_{\epsilon\rightarrow
0}(\overline{z}_1-z_2)=\lim_{\epsilon\rightarrow 0}(t_1-\epsilon
x_1-t_2-\epsilon x_2)=t_1-t_2
\end{eqnarray}
From above equations we obtain
\begin{equation}\label{a}
\lim_{\epsilon\rightarrow
0}G(z_1,z_2,\overline{z}_1,\overline{z}_2)=\delta_{\Delta_1,\Delta_2}\tau^{-2\Delta}
\end{equation}
which is agree with our result (\ref{21}). (For scalar field $\xi_i$
is equal to zero.)\\
Now using above method,  we calculate three-point correlation
function in semi-infinite space-time with a boundary condition at
surface  $x=0$.\\
 Consider the three-point function as
\begin{eqnarray}\label{}
G(x_1,x_2,x_3,t_1,t_2,t_3)=<\phi_1(x_1,t_1)\phi_2(x_2,t_2)\phi_3(x_3,t_3)>
\end{eqnarray}
where $\phi_1,\phi_2$ and $\phi_3$ are Galilean conformal invariant fields.
Invariance  under the time translation symmetry  implies    $G=G(x_1,x_2,x_3,\tau,\sigma)$ where $\tau=t_1-t_3$ and
$\sigma=t_2-t_3$. We constrain $G$ by  scale invariance as
\begin{eqnarray}\label{s1}
<0\mid[L_0,\phi_1\phi_2\phi_3]\mid0>=0~~~~~~~~~~~~~~~~~~~~~~~~~~~~~~~~~~~~~~~~~~\\
\nonumber
\Rightarrow \sum_{i=1}^{i=3}(t_{i}\partial_{t_{i}}+x_{i}\partial_{i}+\Delta_{i})G~~~~~~~~~~~~~~~~~~~~~~~~~~~~~~~~~~~~~~~~~~~~~\\
\nonumber =(\tau
\partial_{\tau}+\sigma\partial_{\sigma}+x_{1}\partial_{x_{1}}+r_{2}\partial_{x_{2}}+x_{3}\partial_{x_{3}}+\Delta_1+\Delta_2+\Delta_3)G=0
\end{eqnarray}
From the invariance under the time component of non-relativistic special
conformal transformation we get
\begin{eqnarray}\label{}
<0\mid[L_1,\phi_1\phi_2\phi_3]\mid0>=0~~~~~~~~~~~~~~~~~~~~~~~~~~~~~~~~~~~~~~~~~~~~~~~~~~~~~~~~~~~~~\\
\nonumber \Rightarrow
\sum_{i=1}^{i=3}(t_{i}^{2}\partial_{t_{i}}+2t_{i}x_{i}\partial_{i}+2t_{i}\Delta_{i}-x_{i}\xi_{i})G~~~~~~~~~~~~~~~~~~~~~~~~~~~~~~~~~~~~~~~~~~~~~~\\
\nonumber
= ((t_{1}^2-t_{3}^2)\partial_{\tau}+(t_2^2-t_3^2)\partial_{\sigma}+2(t_{1}x_{1}\partial_{x_{1}}+t_{2}x_{2}\partial_{x_{2}}+t_3 x_3\partial_3)~~~~~~~~~~~~~~~~~~~~~\\
\nonumber
+2(t_{1}\Delta_{1}+t_{2}\Delta_{2}+t_3\Delta_3)-x_{1}\xi_{1}-x_{2}\xi_{2}-x_3\xi_3)G~~~~~~~~~~~~~~~~~~~~~~~~~~~~~~~~~~\\
\nonumber
=(\tau^2\partial_{\tau}+\sigma^2\partial_{\sigma}+2t_{3}(\tau\partial_{\tau}+\sigma\partial_{\sigma}+x_{1}\partial_{x_{1}}+x_{2}\partial_{x_{2}}+x_3\partial_3)~~~~~~~~~~~~~~~~~~~~~~~\\
\nonumber -x_1\xi_1-r_2\xi_1-x_3\xi_3+2\tau
x_1\partial_{x_{1}}+2\sigma x_2 \partial_{x_2}
+2(t_1\Delta_1+t_2\Delta_2+t_3\Delta_3))G~~~~~~~\\
\nonumber =(\tau^2\partial_{\tau}-x_1\xi_1-x_2\xi_2-r_3\xi_3+2\tau
x_1\partial_{x_{1}}~~~~~~~~~~~~~~~~~~~~~~~~~~~~~~~~~~~~~~~~~~~~~\\
\nonumber +\sigma^2\partial_{\sigma}+2\sigma x_2\partial_{x_2}
+2\tau\Delta_{1}+2\sigma\Delta_2)G=0~~~~~~~~~~~~~~~~~~~~~~~~~~~~~~~~~~~~~~~~~~~
\end{eqnarray}
where in the last equation we have used  Eq.(\ref{s1}). We make the ansatz
\begin{eqnarray}\label{}
G=\delta_{\Delta_1+\Delta_2,\Delta_3}\tau^{-2\Delta_1}\sigma^{-2\Delta_2}G'
\end{eqnarray}
and  simplify  the above equations as
\begin{eqnarray}\label{}
(\tau^2\partial_{\tau}+2\tau
x_1\partial_{x_1}-x_1\xi_1-x_3\xi_3)G'_1=0\\
\nonumber
(\tau\partial_{\tau}+x_1\partial_{x_1}+x_3\partial_{x_3})G'_1=0~~~~~~~~~~~~\\
\nonumber (\sigma^2+2\sigma x_2\partial_{x_2}-x_2\xi_2)G'_2=0~~~~~~~~~~~~\\
\nonumber
(\sigma\partial_{\sigma}+x_2\partial_{x_2})G'_2=0~~~~~~~~~~~~~~~~~~~~~~
\end{eqnarray}
where $G'=G'_1(x_1,x_3,\tau)G'_2(x_2,\sigma)$,
or
\begin{eqnarray}\label{}
(\sigma^2\partial_{\sigma}+2\sigma
x_2\partial_{x_2}-x_2\xi_2-x_3\xi_3)G'_1=0~~~\\
\nonumber
(\sigma\partial_{\sigma}+x_2\partial_{x_2}+x_3\partial_{x_3})G'_1=0~~~~~~~~~~~~~~\\
\nonumber (\tau^2+2\tau x_1\partial_{x_1}-x_1\xi_1)G'_2=0~~~~~~~~~~~~~~\\
\nonumber
(\tau\partial_{\tau}+x_1\partial_{x_1})G'_2=0~~~~~~~~~~~~~~~~~~~~~~~~
\end{eqnarray}
where $G'=G'_1(x_2,x_3,\sigma)G'_2(x_1,\tau)$.
By using the method of characteristic \cite{ka}, we may found the general solution of these equations
\begin{eqnarray}\label{}
G=\delta_{\Delta_1+\Delta_2,\Delta_3}(t_1-t_3)^{-2\Delta_1}(t_2-t_3)^{-2\Delta_2}\exp(\frac{x_1\xi_1}{t_1-t_3}+\frac{x_2\xi_2}{t_2-t_3})\\
\nonumber(\chi_{1}(\frac{x_1x_3}{(t_1-t_3)^2})\exp(-\frac{x_3\xi_3}{t_1-t_3})+\chi_2(\frac{x_2x_3}{(t_2-t_3)^2})\exp(-\frac{x_3\xi_3}{t_2-t_3}))\\
\nonumber +\Sigma (exchanging~~  ~~ 2\leftrightarrow 3 ~~or~~
1\leftrightarrow 3)~~~~~~~~~~~~~~~~~~~~~~~~~~~~~~
\end{eqnarray}
where $\chi_1$ and $\chi_2$ are arbitrary functions. Three-point
function near the boundary is obviously  different from other places.
\section{Holographic dual of non-relativistic BCFT  }
Recently, holographic dual of BCFT was  considered
\cite{ta,fu,al}. $AdS_3$ with Neumann boundary condition at
surface $z=\overline{z}$ is holographic dual of $BCFT_2$. In this
situation the symmetry group of boundary conformal field theory
is generated by one copy of Virasoro algebra \cite{al}. We
introduce non-relativistic version of this gravity dual as a
holographic dual of non-relativistic  BCFT. The $AdS_3$ metric in
Poincare coordinates is
\begin{equation}\label{23}
ds^2=\frac{1}{r^2}(-dt^2+dr^2+dx^2)
\end{equation}
where $r$ is a radial coordinate and $(x,t)$ are boundary
coordinate. In the Eddington-Finkelstein coordinates which define
by $r=r'$ and $t=t'+r'$ the $AdS_3$ metric is given by
\begin{equation}\label{b}
ds^2=\frac{1}{r^2}(dt^2-2dtdr+dx^2)
\end{equation}
The Killing vectors of $AdS_3$ read as
\begin{eqnarray}\label{24}
P=\partial_x~~~~~~~~~~~~~~~~~~~~~~~~B=(t-r)\partial_x-x\partial_{t}~~~~\\
\nonumber
K_x=(t^2-2tr-x^2)\partial_x+2tx\partial_t+2rx\partial_r+2x^2\partial_x\\
\nonumber
H=-\partial_t~~~~~~~~~~~~~~~~~~~~D=-t\partial_t-r\partial_r-x\partial_x~~~\\
\nonumber
K=-(t^2+x^2)\partial_t-2r(t-r)\partial_r-2(t-r)x\partial_x~~~~
\end{eqnarray}
In non-relativistic limit
\begin{equation}\label{25}
t\rightarrow t~~~~~~~r\rightarrow r~~~~~~~x\rightarrow\epsilon x
\end{equation}
we obtain the contracted Killing vectors
\begin{eqnarray}\label{26}
P=\partial_x~~~~B=(t-r)\partial_x~~~~~~K_x=(t^2-2tr)\partial_x~~~~H=-\partial_t~~\\
\nonumber
D=-t\partial_t-r\partial_r-x\partial_x~~~~~~~K=-t^2\partial_t-2(t-r)(r\partial_r+x\partial_x)
\end{eqnarray}
We can define an infinite extension of these vector fields in the
bulk \cite{co}
\begin{eqnarray}\label{27}
M^{(n)}=(t^{n+1}-(n+1)rt^n)\partial_x~~~~~~~~~~~~~~~~~~~~~~~~~~\\
\nonumber
L^{(n)}=-t^{n+1}\partial_t-(n+1)(t^n-nrt^{n-1})(x\partial_x+r\partial_r)
\end{eqnarray}
These vector fields obey the commutation relation (\ref{4})
(without central charges). We can see that these vector fields
at the boundary $r=0$ reduce to Killing vectors of contracted
$CFT_2$ (\ref{3}). The vector fields $M^{(n)}$ only act on the
spatial coordinate $x$, so if we have a boundary condition at
surface $x=0$ ($z=\overline{z}$), these vector fields are removed
from all Killing vectors in the bulk. Now we consider the action
of the  Virasoro generators $L^{(n)}$ (remanded Killing vectors)
on $AdS_3$ metric (\ref{b}) in non-relativistic limit. We
introduce non-relativistic limit of $AdS_3$ metric which is given
by $AdS_2\times R$ metric \cite{co}
\begin{equation}\label{28}
ds^2=\frac{1}{r^2}(-2dtdr+dt^2+dx^2)\rightarrow\frac{1}{r^2}(-2dtdr+dt^2)
\end{equation}
The components of the metric in the ($t,r$) directions survive
and we receive to $AdS_2$ metric. The spatial direction $x$ is
fiber over this $AdS_2$. Virasoro  generators $L^{(n)}$ act
non-trivially on all coordinate
\begin{eqnarray}\label{29}
r\rightarrow r{'}=r(1+a_n(n+1)(t^n-nrt^{n-1}))\\
\nonumber t\rightarrow t{'}=t(1+a_nt^n)~~~~~~~~~~~~~~~~~~~~~~~~~
\end{eqnarray}
where $a_n$ is infinitesimal parameter. From above equation we
have
\begin{eqnarray}\label{30}
dr\rightarrow dr'=dr(1+a_n(n+1)(t^n-nrt^{n-1}))~~~~~~~~~~~~\\
\nonumber
+ra_nn(n+1)t^{n-2}((t-(n-1)r)dt-tdr)\\
\nonumber dt\rightarrow
dt'=dt(1+(n+1)a_nt^n)~~~~~~~~~~~~~~~~~~~~~~~~~~~
\end{eqnarray}
So in non-relativistic limit (\ref{25}) we get
\begin{equation}\label{31}
ds^2=\frac{1}{r^2}(-2dtdr+dt^2)\rightarrow
\frac{1}{r^2}(-2dtdr+dt^2+2n(n^2-1)a_nr^2t^{n-2}dt^2)
\end{equation}
The $SL(2,R)$ subgroup which is generated by $L^0,L^{\pm}$ are
exact isometries of non-relativistic version of boundary $AdS_3$.
Near the boundary $r=0$ the diffeomorphisms  of above metric has
a fall-off like $r^2$, so other $L^{n}$ are asymptotic isometries
of non-relativistic $AdS_3$. One copy of Virasoro algebra is
asymptotic symmetry of non-relativistic $AdS_3$ metric with a
boundary condition. Following \cite{va}, we calculate two-point
function from gravity dual. Equation of motion for massive scalar
field on the $AdS_3$ background (\ref{23}) is given by
\begin{equation}\label{32}
\frac{1}{\sqrt{G}}\partial_{M}(\sqrt{G}G^{MN}\partial_N\phi(t,r,x))-m^2\phi(t,r,x)=0
\end{equation}
In non-relativistic limit (\ref{25}) we have
\begin{equation}\label{33}
\frac{1}{\sqrt{G}}\partial_{a}(\sqrt{G}g^{ab}\partial_b\phi)-m^2\phi=0~~~~~~~\partial_x^2\phi=0
\end{equation}
The first equation can be obtained from the following action
\begin{equation}\label{34}
I=\int dt dr
\sqrt{G}\frac{1}{2}(g^{ab}\partial_a\phi\partial_b\phi+m^2\phi^2)
\end{equation}
General solution of the equation of motion of the above action is
\begin{equation}\label{35}
\phi(t,r)=r e^{-i\omega t}(AI_{\alpha}(\omega r)+B
K_{\alpha}(\omega r))
\end{equation}
where $\alpha=\sqrt{m^2+1}$. From $AdS_3/CFT_2$ correspondence,
we can find the bulk solution as
\begin{equation}\label{36}
\phi(t,r)=c\delta^{\Delta-2}\int dt'
\phi_{\delta}(t')(\frac{r}{r^2+|t-t'|^2})^{\Delta}
\end{equation}
where $\phi_{\delta}$ is a Dirichlet boundary value at $r=\delta$
and $\Delta=\alpha+\frac{1}{2}$. The above equation can be used
to read two-point function of $GCA_2$
\begin{equation}\label{37}
<\phi(t_1)\phi(t_2)>\sim (t_1-t_2)^{-2\Delta}
\end{equation}
which agrees with  results (\ref{21}) and (\ref{a}).
\section{Conclusion}
Galilean conformal algebra (GCA) arises as a contraction of
conformal algebra. We can use 2d GCA to constrain correlation
functions. Correlation functions of Galilean conformal invariant
fields in 2d for space-time without boundary condition were found
in \cite{ba}. We calculated two-point and three-point functions in semi-infinite
space with a boundary condition at surface $z=\overline{z}$
($x=0$), by using  some methods in quantum field theory
(\ref{21}) and from gravity dual (\ref{37}). Our results
(\ref{21}) and (\ref{37}) are agree with two-point function of
BCFT in non-relativistic limit (\ref{a}). We also introduce
holographic dual of BCFT in non-relativistic limit (BGCA).
$AdS_3$ with boundary condition and in non-relativistic limit has
asymptotic isometries which are generated by one copy of
non-relativistic version of  Virasoro algebra .

\end{document}